\documentclass[aps,prl,twocolumn,groupedaddress,amsmath,amssymb]{revtex4}

\usepackage{graphicx}% Include figure files
\usepackage{dcolumn}% Align table columns on decimal point
\usepackage{bm}% bold math

\begin{document}

\title{Giant overlap between the magnetic and superconducting phases of CeAu$_{2}$Si$_{2}$ under pressure}

\date{\today}
\author{Z. Ren$^{1}$, L. V. Pourovskii$^{2,3}$, G. Giriat$^{1}$, G. Lapertot$^{4,5}$, A. Georges$^{2,6,1}$ and D. Jaccard$^{1}$}
\email{Didier.Jaccard@unige.ch}
\affiliation{$^{1}$DPMC - University of Geneva, 24 Quai Ernest-Ansermet, 1211 Geneva 4, Switzerland}
\affiliation{$^{2}$Centre de Physique Th$\acute{e}$orique, CNRS, $\acute{E}$cole Polytechnique, 91128 Palaiseau, France}
\affiliation{$^{3}$Swedish e-science Research Centre (SeRC), Department of Physics, Chemistry and Biology (IFM), Link$\ddot{o}$ping University, Link$\ddot{o}$ping, Sweden}
\affiliation{$^{4}$University of Grenoble Alpes, INAC-SPSMS, F-38000 Grenoble, France}
\affiliation{$^{5}$CEA, INAC-SPSMS, F-38000 Grenoble, France}
\affiliation{$^{6}$Coll$\grave{e}$ge de France, 11 place Marcelin Berthelot, 75005 Paris, France}

\begin{abstract}
High pressure provides a powerful means for exploring unconventional superconductivity which appears mostly on the border of magnetism. Here we report the discovery of pressure-induced heavy fermion superconductivity up to 2.5 K in the antiferromanget CeAu$_{2}$Si$_{2}$ ($T_{\rm N}$ $\approx$ 10 K). Remarkably, the magnetic and superconducting phases are found to overlap across an unprecedentedly wide pressure interval from 11.8 to 22.3 GPa. Moreover, both the bulk $T_{\rm c}$ and $T_{\rm M}$ are strongly enhanced when increasing the pressure from 16.7 to 20.2 GPa. $T_{\rm c}$ reaches a maximum at a pressure slightly below $p_{\rm c}$ $\approx$ 22.5 GPa, at which magnetic order disappears. Furthermore, the scaling behavior of the resistivity provides evidence for a continuous delocalization of the Ce 4$f$-electrons associated with a critical endpoint lying just above $p_{\rm c}$. We show that the maximum $T_{\rm c}$ of CeAu$_{2}$Si$_{2}$ actually occurs at almost the same unit-cell volume as that of CeCu$_{2}$Si$_{2}$ and CeCu$_{2}$Ge$_{2}$, and when the Kondo and crystal field splitting energies becomes comparable. Dynamical mean-filed theory calculations suggest that the peculiar behavior in pressurized CeAu$_{2}$Si$_{2}$ might be related to its Ce 4$f$-orbital occupancy. Our results not only provide a unique example of the interplay between superconductivity and magnetism, but also underline the role of orbital physics in understanding Ce-based heavy fermion systems.
\end{abstract}

% on superconducting transition temperature, 74.62.Fj
% High-pressure effects in solids and liquids, 62.50.-p
% noncuprate superconductors, 74.70.Tx

\maketitle
\section{I. Introduction}
Superconductivity and magnetism were long thought to be antagonistic. In this context, the discovery 22 years ago that high pressure turns a magnetically ordered heavy fermion (HF) compound, namely CeCu$_{2}$Ge$_{2}$, into a superconductor has attracted much attention in the condensed matter physics community \cite{CeCu2Ge2Jaccard}. Since then, high pressure studies of Ce-based HFs have revealed a number of superconductors such as CePd$_{2}$Si$_{2}$ \cite{CePd2Si2}, CeIn$_{3}$ \cite{CeIn3}, CeRhIn$_{5}$ \cite{CeRhIn5} and, more recently, CePt$_{2}$In$_{7}$ \cite{CePt2In7}. In all known cases, SC emerges in the vicinity of a magnetic-nonmagnetic phase boundary, and most often competes for stability with magnetic order, except in a few examples where both states coexist within a narrow pressure range \cite{ReviewKnebel}.

CeCu$_{2}$Ge$_{2}$ is an isostructural sister compound of the first discovered HF superconductor CeCu$_{2}$Si$_{2}$ \cite{CeCu2Si2steglich}, which exhibits a second superconducting phase under pressure with a higher $T_{\rm c}$ \cite{CeCu2Si284,ReviewDidier,CeCu2Si2Holmes,ReviewJaccard2,CeCu2Si2seyfarth}. Remarkably, CeCu$_{2}$Ge$_{2}$ shares the same phase diagram as CeCu$_{2}$Si$_{2}$ when pressurized above $\sim$ 10 GPa where magnetism disappears \cite{vargoz,ReviewDidier}. Indeed, both compounds feature the existence of connected low- and high-pressure superconducting phases associated with two critical points of different origins. Moreover, the partial substitution of Si by Ge in CeCu$_{2}$Si$_{2}$ results in the splitting of the initially joined superconducting phases into two distinct domes due to disorder-induced pair breaking \cite{CeCu2Si2yuan}.

Despite three decades of efforts, the underlying mechanisms for the two superconducting phases are still poorly understood. On one hand, it is widely believed that SC at low pressure is mediated in these systems by critical spin fluctuations \cite{CeCu2Si2stockert,Lonzarich}. However, this view was very recently challenged by a thermodynamical study which points to multiband SC with a full energy gap in CeCu$_{2}$Si$_{2}$ at ambient pressure \cite{CeCu2Si2multiband}. On the other hand, there is no general consensus that critical valence fluctuations are responsible for high pressure SC \cite{CeCu2Si2Holmes,CeCu2Si2XAS}. An alternative interpretation is that critical fluctuations stemming from orbital transition provide the glue of the superconducting pairing \cite{Hattori2010,Pourovskii2014}.

In this regard, the exploration of high pressure SC in close relatives of CeCu$_{2}$Si$_{2}$ is highly desirable. For such investigation the isoelectronic and isostructural compound CeAu$_{2}$Si$_{2}$ is an excellent candidate. At ambient pressure, CeAu$_{2}$Si$_{2}$ orders antiferromagnetically below $T_{\rm N}$ $\approx$ 10 K. A previous high-pressure study of a polycrystalline sample, carried out down to 1.2 K, shows that while the magnetic order disappears around 16 GPa, SC does not occur below 19.5 GPa \cite{CeAu2Si2Link}.

In this paper, we report on pressure-induced SC in high quality CeAu$_{2}$Si$_{2}$ single crystals with $T_{\rm c}$ reaching 2.5 K observed from high-pressure $"$multiprobe$"$ (transport and calorimetry) measurements up to 27.4 GPa. Unexpectedly, the resulting pressure-temperature phase diagram reveals a highly unusual interplay of superconductivity with magnetism, and differs markedly from that of all known Ce-based pressure induced superconductors. In particular, for the first time, both superconductivity and magnetism are enhanced with increasing pressure over a broad pressure region. We present a comparison of the unit-cell volume phase diagram of CeAu$_{2}$Si$_{2}$ with that of CeCu$_{2}$Si$_{2}$ and CeCu$_{2}$Ge$_{2}$, and discuss the implication of these results on the pairing mechanism. First-principle calculations show the existence of an intermediate state in the pressure dependence of the Ce 4$f$-orbital occupancy for CeAu$_{2}$Si$_{2}$, which may be a key ingredient for understanding the peculiar behavior in this compound.

\section{II. Methods}
Single crystals of CeAu$_{2}$Si$_{2}$ were grown by the "flux" method (see Ref. \cite{canfield} for guide lines) using Au-Si self flux and Sn flux. The starting materials were Ce (99.99\%) from Ames Lab \cite{elementCe}, Au (99.999\%), Si (99.9999\%), and Sn (99.9999\%) from Alfa-Aesar. In the self-flux method, elements with a ratio of Ce$_{0.05}$Au$_{0.475}$Si$_{0.475}$ were melted in an alumina crucible inside a sealed evacuated quartz ampoule, held at 1120 $^{\circ}$C for 6 h, followed by a slow cooling at 1.2 $^{\circ}$C/h down to 850 $^{\circ}$C. In the Sn-flux method, which is slightly different from that described in Ref. \cite{CeAu2Si2crystalgrowth}, an ingot of Ce:Au:Si = 1:20:3 was pre-synthesized by arc melting under an argon atmosphere, and flipped five times to ensure homogeneity. Big pieces of the crushed ingot were melted with Sn (CeAu$_{20}$Si$_{3}$:Sn = 1:50) in an alumina crucible inside a sealed evacuated quartz ampoule, held at 1150 $^{\circ}$C for 48 h, followed by a slow cooling at 1 $^{\circ}$C/h down to 650 $^{\circ}$C. In both cases, crystals, separated from the flux by centrifugation, exhibit well-developed facets and have sizes of up to a few cubic millimeters. Phase impurities were not detected neither by XRD nor by SEM-EDX measurements. The ambient-pressure measurements show that, within the uncertainty of the geometrical factor, the difference between the Sn- and the self-flux samples' resistivity values is almost temperature independent, and their respective residual resistivities are 1.8 and 12.2 $\mu$$\Omega$cm.

High pressure experiments were performed in a Bridgman-type sintered diamond anvil pressure-cell using steatite as pressure transmitting medium and lead (Pb) as pressure gauge \cite{Holmesiron}.
A photograph of the actual setup is shown in Fig. S1 of the Supplemental Material \cite{SM}.
The CeAu$_{2}$Si$_{2}$ sample, arranged in such a way that the $c$-axis is parallel to the compressive force, is connected in series with the Pb for four-probe resistivity measurements.
The ac-calorimetry measurements and data analysis were carried out according to the method described in Ref. \cite{Holmesthesis}.
The chromel wire, which otherwise served as one of the voltage leads, was used as the heater and was thermally excited by an ac current of frequency $\omega$ while the sample temperature oscillations were detected by measuring the voltage of the Au/AuFe thermocouple $V_{\rm ac}$ at a frequency of 2$\omega$. The data recorded at frequencies above and well below the cutoff frequency correspond respectively to a signal dominated by the sample contribution, which can be considered to be inversely proportional to the heat capacity, and to a measure of the mean elevation of the sample temperature over that of the bath. Note that the resistance of the chromel wire is almost pressure independent and more than two orders of magnitude larger than that of the sample. Hence the possibility that the observed anomalies are due to the drastic change in heating power can be ruled out. Throughout the experiments, the pressure gradient estimated from the width of the Pb superconducting transition was $\Delta$$p$ $\leqslant$ 0.3 GPa. After depressurisation, the examination of the pressure chamber showed that the distance between the voltage leads has increased by less than 10\%. This, together with the geometrical factor uncertainty as well as the change in the sample volume under pressures, set a 15\% error on the absolute resistivity value.

Theoretical calculations were performed based on a combination of
electronic structure and dynamical mean-field theory (DMFT) methods,
which takes into account both the local atomic physics and correlation
effects in the Ce $f$-shell as well as the physical effects of
hybridization with conduction electrons (renormalization of
crystal-field levels and Kondo screening) ~\cite{Pourovskii2014}. Our
self-consistent (over the charge density) implementation of this method is
that of Refs. \cite{Aichhorn2009} and \cite{Aichhorn2011} and uses the
full-potential linear augmented plane-wave electronic structure
Wien2k~\cite{Wien2k} code. The DMFT quantum impurity problem was solved
with the continuous-time quantum Monte-Carlo (CT-QMC)
method~\cite{Gull2011} based on the TRIQS library \cite{TRIQS} package.

The Wannier orbitals representing Ce 4$f$ states were constructed from
the Kohn-Sham states within the energy range from -12.4 to 5.4 eV  using
the projective approach of Ref.~\cite{Aichhorn2009}. The local Coulomb
interaction between Ce 4$f$ electrons was approximated by a
spherically-symmetric density-density form parameterized by the  Slater
parameter $F_0$=$U$=6.36~eV \cite{LDAestimation} and Hund's rule coupling $J$=0.7~eV used previously in
Ref~\cite{Pourovskii2014}. The fully-localized-limit form of the
double-counting correction term was employed throughout.

We used the eigenstates $|\Gamma\rangle$ of Ce$^{3+}$, obtained by
diagonalizing the {\it ab initio} crystal-field and spin-orbit
Hamiltonian, as a basis for CT-QMC calculations. Rather small off-diagonal elements of
the DMFT hybridization function in the $\{\Gamma\}$ basis were
neglected. This approximation allowed us to treat the full Ce 4$f$ shell and to access low temperatures using the fast "segment picture"
algorithm of CT-QMC method \cite{Gull2011}. The DMFT quantum impurity problem
was solved using up to 4$\times$10$^{11}$ CT-QMC moves with a
measurement performed after each 200 moves. The calculations were performed for the body-centered tetragonal
ThCr$_2$Si$_2$-type structure at the experimental value of the lattice
parameters determined as a function of pressure in Refs.
\cite{Onodera2002,CeCu2Ge2latticeparameter,CeAu2Si2latticeparameter}.

\section{III. Results and Discussion}
\subsection{A. Experimental results of CeAu$_{2}$Si$_{2}$}
The general trend of the electrical resistivity of the investigated CeAu$_{2}$Si$_{2}$ single-crystals as a function of pressure ($p$) and temperature ($T$) (see Fig. S2 of the Supplemental Material \cite{SM}) is typical of Ce-based Kondo lattice systems \cite{ReviewDidier}. However, our samples show a thirty times lower residual resistivity ($\sim$ 1.8 $\mu\Omega$cm at $p$ = 0) than reported previously \cite{CeAu2Si2Link}. The temperature dependence of the magnetic contribution $\rho_{\rm mag}$ to resistivity at representative pressures, shown in Fig. 1a, is obtained by subtracting the phonon contribution approximated as a pressure-independent term $\rho_{\rm ph}\approx$ 0.062$\times$\emph{T} ($\mu\Omega$cm) \cite{phononcontribution} from the raw data. Below room temperature, the ambient pressure curve already unveils two anomalies of weak magnitude. One can discern a maximum at $\sim$ 140 K, as well as the onset of a low-temperature anomaly which is masked by the sharp drop in resistivity ascribed to the magnetic ordering below $T_{\rm N}$ $\approx$ 10 K. At an intermediate pressure of 14.2 GPa, these anomalies have grown markedly and $\rho_{\rm mag}$($T$) exhibits two characteristic maxima at temperatures $T_{1}^{\rm max}$ $\approx$ 8 K and $T_{2}^{\rm max}$ $\approx$ 137 K. Above each of these maxima, $\rho_{\rm mag}$($T$) follows specific $-$ ln$T$ dependencies, which reflect the incoherent Kondo scattering of the ground state and excited crystal-field levels, respectively \cite{scattering1}. When increasing pressure up to 20.2 GPa, the Kondo scattering keeps increasing. The temperature $T_{2}^{\rm max}$ remains almost unchanged, while $T_{1}^{\rm max}$ has by now started its rise. At the highest measured pressure of 27.4 GPa, the two maxima have already merged. The contribution at $T_{1}^{\rm max}$ dominates in such a way that a single peak is observed at $\sim$ 200 K. Connected to this behavior, the $-$ ln$T$ resistivity slope becomes steeper with increasing pressure, which is interpreted as resulting from a dramatic rise in Kondo temperature of two and a half orders of magnitude over the investigated pressure range \cite{scattering1}.
\begin{figure}
\centering
\includegraphics*[width=8.5cm]{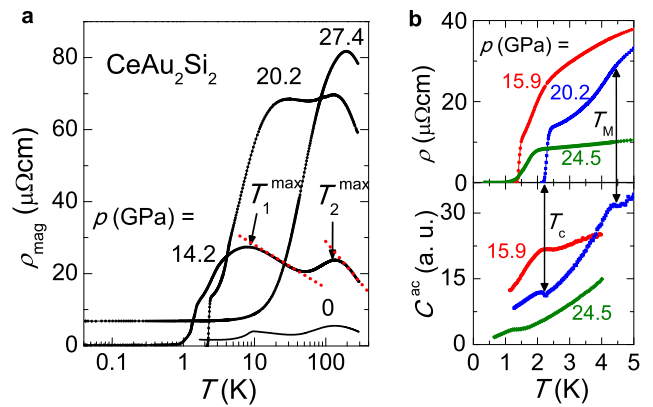}
\caption{(a) Logarithmic temperature dependence of the magnetic contribution $\rho_{\rm mag}$ to the in-plane resistivity of CeAu$_{2}$Si$_{2}$ for typical pressures. The two characteristic maxima $T_{1}^{\rm max}$ and $T_{2}^{\rm max}$ at 14.2 GPa are marked by the arrows. The dotted red lines are a guide to the eyes, showing the -- ln$T$ dependence of the resistivity. Note that with increasing pressure, $T_{1}^{\rm max}$ increases drastically while $T_{2}^{\rm max}$ remains almost constant, and finally the two maxima merge into a single peak.
(b) Comparison of the resistivity $\rho$ and heat capacity $C^{\rm ac}$ for three different pressures at which complete resistive superconducting transitions were observed. A jump in $C^{\rm ac}$ due to SC coincides with the completeness of the resistive transition at 20.2 and 24.5 GPa, but is absent at 15.9 GPa. At 15.9 and 20.2 GPa, the jump in $C^{\rm ac}$ above $T_{\rm c}$ accompanied by a change in the slope of $\rho$ indicates the magnetic ordering. As an example, two arrows show that at 20.2 GPa the jumps in $C^{\rm ac}$ correspond well to the magnetic transition ($T_{\rm M}$ $\sim$ 4.5 K) and to the completeness of the resistive transition ($T_{\rm c}$$^{\rm \rho=0}$ $\sim$ 2.2 K).
}
\label{fig1}
\end{figure}

For the first time, signatures of both superconducting and magnetic transitions were observed in resistivity and heat capacity measurements of CeAu$_{2}$Si$_{2}$, as exemplified in Fig. 1b (see also Fig. S3 of the Supplemental Material \cite{SM}). It can be seen that at 20.2 GPa, the resistive superconducting transition around 2.2 K coincides with a jump ($\sim$ 20\% of the total signal) in $C^{\rm ac}$($T$), indicating bulk SC. Such an anomaly due to SC is detected at pressures as high as 24.5 GPa. By contrast, at 15.9 GPa, despite the sharp and complete resistive transition, no corresponding anomaly is detected in $C^{\rm ac}$($T$), indicating that SC is likely filamentary or textured \cite{texturedSC}. It is worth noting that at 15.9 and 20.2 GPa, the jump in heat capacity together with the downward change of slope in resistivity (distinctively above $T_{\rm c}$), are evidence of the persistence of a magnetic ordering, presumably of antiferromagnetic nature. In line with common practice \cite{determinationofTM}, the superconducting and magnetic ordering temperatures $T_{\rm c}$ and $T_{\rm M}$ in $C^{\rm ac}$($T$) are defined by the midpoint of the respective jumps when considering entropy conservation. As an example, the two up down arrows show that at 20.2 GPa the two midpoints agree well with the completeness of the resistive transition ($T_{\rm c}$$^{\rm \rho=0}$ $\sim$ 2.2 K) and the onset of downward curvature in the resistivity ($T_{\rm M}$ $\sim$ 4.5 K), respectively.

The pressure dependencies of both the $T_{\rm M}$ and $T_{\rm c}$ define the phase diagram shown in Fig. 2a, which reveals several remarkable new features. Even though the magnetism persists up to $\sim$ 22 GPa, SC is found over the very broad pressure range 11.8 $-$ 26.5 GPa, resulting in a giant overlap of the two phases. At low pressure $T_{\rm M}$ first displays a linear in-$T$ decrease due to the increase of the Kondo coupling, in agreement with previous data \cite{CeAu2Si2Link}, but at higher pressures $T_{\rm M}$ shows an atypical nonmonotonic dependence, whose anomalies are clearly connected to SC. Indeed, the emergence of filamentary and bulk SC each correspond to a strengthening of magnetism, which manifests itself in the $T_{\rm M}$ evolution by a flattening and a cusp, respectively. These anomalies may signal the formation of new magnetic phases \cite{CeAuSb2}. In the pressure interval 11.8$-$15.9 GPa, where $T_{\rm c}$ onset increases while $T_{\rm M}$ decreases, SC appears to compete with magnetism as usually observed \cite{ReviewKnebel}. However, at higher pressures up to 20.2 GPa, there is a dramatic rise of the $T_{\rm M}$ from $\sim$ 2.5 to $\sim$ 4.5 K. Remarkably, this rise corresponds to a reduction of the heat capacity anomaly. Moreover, in the same pressure range, the bulk $T_{\rm c}$ also increases strongly (by a similar factor) from $\sim$ 1.4 to $\sim$ 2.3 K while the width of the resistive transition remains narrow ($\Delta$$T_{\rm c}$ $<$ 0.3 K). Consequently, it is unlikely that SC and magnetic order originate from separated phases. Such a simultaneous enhancement of both $T_{\rm M}$ and $T_{\rm c}$ over a broad pressure range has never been reported in any other Ce-based pressure-induced superconductors.
\begin{figure}
\centering
\includegraphics*[width=8cm]{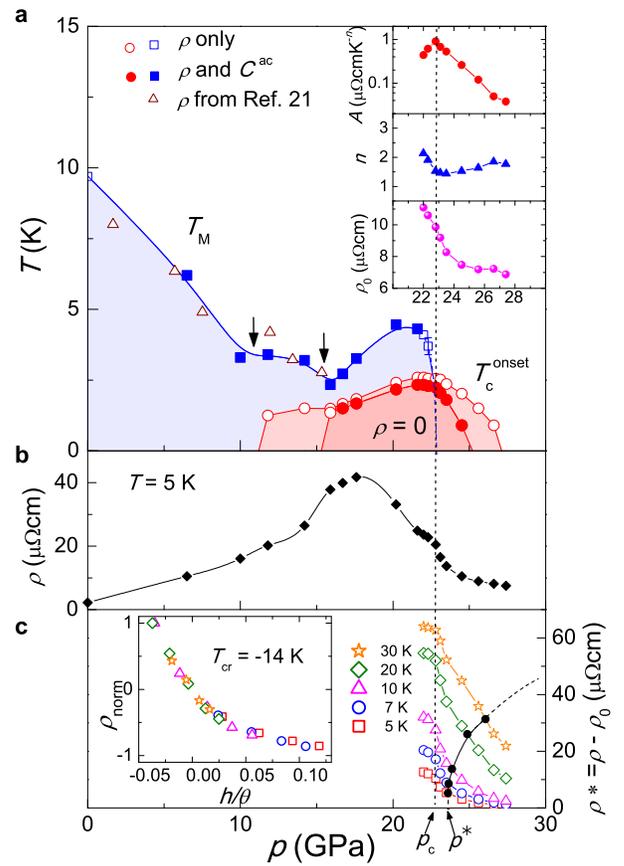}
\caption{(a) Experimental $p$$-$$T$ phase diagram of CeAu$_{2}$Si$_{2}$. $T_{\rm c}^{\rm onset}$ and $T_{\rm M}$ represent the superconducting transition onset and the magnetic ordering temperatures, respectively. The open (closed) symbols denote the data extracted from only the resistivity measurements (from both resistivity and heat capacity measurements). The data from Ref. \cite{CeAu2Si2Link} are also included for comparison, and a good agreement is found. The two arrows mark the anomalies in the $T_{\rm M}$($p$) curve at $\sim$ 11 and $\sim$ 16 GPa, in accordance with the emergence of filamentary and bulk SC. The inset shows the fitting parameters of the power law $\rho(T)$ = $\rho_{0}$ + $AT^{n}$ to the resistivity data above $T_{\rm c}$ plotted as a function of pressure. The $A$ coefficient reaches a maximum and the temperature exponent $n$ exhibits a minimum at $p_{\rm c}$ $\approx$ 22.5 GPa, where the magnetic order disappears.
(b) Plot of $\rho$ versus $p$ at 5 K. The data exhibit one peak at $\sim$ 17 GPa (onset of bulk SC) and a shoulder at $\sim$ $p_{\rm c}$.
(c) Plot of $\rho^{*}$ = $\rho$ -- $\rho_{0}$ versus $p$ at selected temperatures up to 30 K. The solid circles indicate for each isotherm the 50\% drop of $\rho^{*}$ compared to its value at 22.8 GPa.
The inset shows the collapse of all normalized data $\rho_{\rm norm}$ when plotted against the generalized distance $h/\theta$ from the critical end point located at $p^{*}$ $\approx$ 23.6 GPa and $T_{\rm cr}$ = --14 K (see text for details).
The two critical pressures $p_{\rm c}$ and $p^{*}$ are indicated by labeled arrows.
}
\label{fig2}
\end{figure}

With increasing pressure above 21.6 GPa, the signature of magnetic ordering is present only in resistivity, and $T_{\rm M}$ (see Fig. S4 of the Supplemental Material \cite{SM} for its determination) decreases rapidly and drops to zero between 22.3 and 22.8 GPa, indicating a magnetic quantum critical point (QCP) at $p_{\rm c}$ = 22.5 $\pm$ 0.3 GPa. Correspondingly, the fitting of the power law $\rho(T)$ = $\rho_{0}$ + $AT^{n}$ to the resistivity data above $T_{\rm c}$, where $\rho_{0}$ is the residual resistivity, yield a maximum $A$ coefficient and a minimum exponent ($n$ $\approx$ 1.5) at $p_{\rm c}$, both of them being standard signatures of a QCP (see Inset Fig.2a). Note that the uncertainty in $p_{\rm c}$ is comparable to the pressure gradient inside the pressure chamber, and thus the phase transition could be weakly first order, which raises questions about the quantum nature of $p_{\rm c}$. On the other hand, at pressures below 22 GPa, the power law analysis is not pertinent due to the strong magnetic contribution to the resistivity below $T_{\rm M}$. In order to highlight the ground state excitations independently of magnetic ordering, we plot the isothermal resistivity at 5 K, $i.e.$ just above $T_{\rm M}$ for $p$ $>$ 7 GPa, versus pressure, as shown in Fig. 2b. Clearly the resistivity at 5 K shows a broad peak of high magnitude at around 17 GPa, which roughly coincides with the local minimum in $T_{\rm M}$ and the onset of bulk SC. This observation supports the existence of a putative QCP around the pressure at which the maximum scattering rate occurs. At $p_{\rm c}$, only a weak anomaly is seen whose magnitude is smaller than the term $AT^{1.5}$ (with $T$ = 5 K), as expected.

Slightly below $p_{\rm c}$, $T_{\rm c}$ reaches a maximum of $\sim$ 2.5 K, which is among the highest value reported to date for Ce-based HF superconductors. In order to further characterize the superconducting state, we have measured the resistive transition at 22.3 GPa under various magnetic fields ($B$) applied along the crystal's $c$-axis (see Fig. S5 of the Supplemental Material \cite{SM}). The results show a very large initial slope of the upper critical field ($dB_{\rm c2}/dT$)$_{T_{\rm c}}$ = $-$7.1 T/K, and given that $|$$dB_{\rm c2}/dT$$|$$_{T_{\rm c}}$ $\propto$ $m^{*2}$$T_{\rm c}$ \cite{Hc2slope}, a very large effective mass $m^{*}$ $\sim$ 110$m_{\rm 0}$ ($m_{\rm 0}$ is the free electron mass) is obtained, confirming heavy fermion SC. Furthermore, using the extrapolated upper critical field at zero temperature $B_{\rm c2}$(0) $\sim$ 9.2 T, the Ginzburg-Landau (GL) coherence length $\xi_{\rm GL}$ is estimated as $\xi_{\rm GL}$ = $\sqrt{\Phi_{0}/2\pi B_{\rm c2}(0)}$ $\approx$ 55 {\AA}, where $\Phi_{0}$ is the flux quantum. Preliminary measurements show that ($dB_{\rm c2}/dT$)$_{T_{\rm c}}$ scales with $T_{\rm c}$.

Just above the pressure $p_{\rm c}$, in parallel with the decrease of $T_{\rm c}$, the low temperature isothermal resistivity $\rho^{*}$($p$) = $\rho$($p$) $-$ $\rho_{\rm 0}$($p$) goes down steeply with increasing pressure, as shown in Fig. 2c. This behavior is reminiscent of what was found in the vicinity of the second critical point of CeCu$_{2}$Si$_{2}$ around 4.5 GPa, a behavior that was analyzed assuming an underlying critical endpoint located at ($p_{\rm cr}$, $T_{\rm cr}$) in the $p$$-$$T$ plane \cite{CeCu2Si2seyfarth}. Following the same data treatment, we define the normalized resistivity $\rho_{\rm norm}$($p$) = ($\rho^{*}$($p$) $-$ $\rho^{*}$($p_{\rm 50\%}$))/$\rho^{*}$($p_{\rm 50\%}$), where for each temperature, $p_{\rm 50\%}$ denotes the pressure corresponding to the midpoint of the $\rho^{*}$($p$)-drop compared to its value at 22.8 GPa. As seen in the inset of Fig. 2b, the $\rho_{\rm norm}$ data below 30 K collapse onto a single curve, when plotted as a function of $h/\theta$, where $h$ = ($p$ $-$ $p_{\rm 50\%}$)/$p_{\rm 50\%}$ and $\theta$ = ($T$ $-$ $T_{\rm cr}$)/$|T_{\rm cr}|$ with the only free parameter $T_{\rm cr}$ = $-$ 14 (2) K. Such a scaling shows that the resistivity is fully governed by the proximity of a critical endpoint in a broad region of the $p$ $-$ $T$ plane ($p$ $>$ $p_{\rm c}$, $T$ $\leq$ 30 K). Moreover, the slightly negative $T_{\rm cr}$ value substantiates that CeAu$_{2}$Si$_{2}$ just misses a first-order transition, meaning that only a
crossover occurs. By extrapolating the temperature dependence of $p_{\rm 50\%}$ to zero temperature, we obtained $p^{*}$ ($\approx$ $p_{\rm cr}$) = 23.6 $\pm$ 0.5 GPa.

To check the reproducibility of the above results, we performed measurements on CeAu$_{2}$Si$_{2}$ crystals, grown by a self-flux (Au-Si) method (unpublished results). Although the residual resistivity $\rho_{\rm 0}$ of these crystals is about five times higher than that of the present study, the pressures $p_{\rm c}$ and $p^{*}$ were found to be almost identical to the aforementioned values, clearly indicating that they are intrinsic and not affected by the sample quality. Around $p^{*}$, a scaling of resistivity was also obtained with a slightly more negative $T_{\rm cr}$. Moreover, we observed a similar resurgence of magnetism for $p$ $>$ 15 GPa. However, for the self-flux grown crystals, SC emerges only from 20 GPa and the maximum $T_{\rm c}$ $\sim$ 1.1 K is considerably lower. We attribute this to a strong pair breaking effect, especially in the magnetic phase, when the electron mean free path $\ell$ $\propto$ 1/$\rho_{\rm 0}$ is short, consistent with observations in CeCu$_{2}$Ge$_{2}$ \cite{ReviewDidier}. This also explains why no SC was detected at all in the previous study which was performed on polycrystalline samples with an even higher $\rho_{\rm 0}$ value \cite{CeAu2Si2Link}.

\subsection{B. Comparison with CeCu$_{2}$Si$_{2}$ and CeCu$_{2}$Ge$_{2}$}
In order to allow for a straightforward comparison of CeAu$_{2}$Si$_{2}$ and CeCu$_{2}$$X$$_{2}$ ($X$ = Si or Ge), we have converted the pressure into the unit-cell volume ($V$) of each compound, using high-pressure crystallographic results \cite{Onodera2002,CeCu2Ge2latticeparameter,CeAu2Si2latticeparameter} (see Fig. 3 caption). The three corresponding $V$ $-$ $T$ phase diagrams are shown in Fig. 3. Strikingly, despite the very different properties observed at ambient pressure (notably the ambient pressure volume $V_{\rm 0}$), there is a broad overlap of the bulk superconducting regions of the three compounds, which confirms that the local environment of the Ce ions plays a key role in the occurrence of SC. In particular, the $V$ dependencies of $T_{\rm c}$ are nearly identical for CeCu$_{2}$$X$$_{2}$ and their maximum $T_{\rm c}$s occur at the same $V^{\ast}$ $\thickapprox$ 158 {\AA}$^{3}$, far away from the volume at which magnetism disappears in CeCu$_{2}$Ge$_{2}$ and in good agreement with a previous report \cite{CeCuSi2NQR}. This excellent match resembles the one obtained for the magnetic phase diagrams of CePd$_{2}$Si$_{2}$ and CePd$_{2}$Ge$_{2}$, when plotted versus their $V$ \cite{CePd2X2}.
\begin{figure}
\includegraphics*[width=8cm]{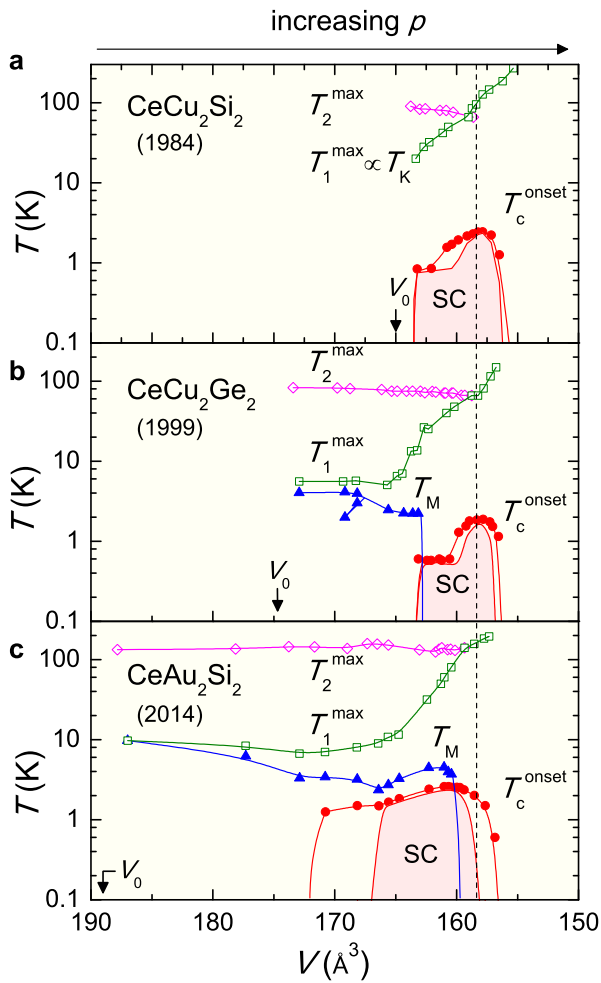}
\caption{(a) CeCu$_{2}$Si$_{2}$,
(b) CeCu$_{2}$Ge$_{2}$, and (c) CeAu$_{2}$Si$_{2}$. Data for CeCu$_{2}$Ge$_{2}$ and CeCu$_{2}$Si$_{2}$ are taken from Ref. \cite{ReviewDidier} and \cite{CeCu2Si2seyfarth}, respectively.
The unit-cell volumes ($V_{\rm 0}$) at ambient pressure and room temperature for each compound are indicated by the arrows. The small temperature dependence (within 1.2 \%) of the unit-cell volume has been taken into account as described in Note 1 of the Supplemental Material \cite{SM}, and thus the data for each compound at ambient pressure is actually located at a volume smaller than $V_{\rm 0}$.
Note that for the three compounds the two resistivity maxima merge ($T_{1}^{\rm max}$ = $T_{2}^{\rm max}$) and also $T_{\rm c}$ reaches its maximum near the same volume of 158 {\AA}$^{3}$, as indicated by the vertical dashed line. In comparison with CeCu$_{2}$$X_{2}$, in CeAu$_{2}$Si$_{2}$ the SC emerges at a larger $V$ ($\sim$ 171 {\AA}$^{3}$) but the magnetic order persists down to a smaller $V$ ($\sim$ 160 {\AA}$^{3}$).
}
\label{fig3}
\end{figure}

For CeAu$_{2}$Si$_{2}$ the maximum $T_{\rm c}$ is slightly shifted to a larger $V$. This shift of about 2 {\AA}$^{3}$ seems beyond experimental error and might indicate the limit of our comparison in terms of the unit-cell volume. Since we compare isostructural and isoelectronic compounds, the unit-cell volume is certainly a relevant parameter, although it is not ideal. For instance, when pressure reduces the $V$ of CeAu$_{2}$Si$_{2}$ to that of CeCu$_{2}$Si$_{2}$, the Au and the Cu atoms may still contribute differently to the crystal field effect.

The phase diagram of CeAu$_{2}$Si$_{2}$ exhibits two qualitative differences with that of CeCu$_{2}$$X$$_{2}$. Firstly, SC emerges deep inside the magnetic region. Secondly, magnetic ordering persists down to a smaller $V$, and its disappearance coincides with the maximum $T_{\rm c}$. This latter characteristic, which is the most common case for Ce-based pressure-induced superconductors, can be taken as strong evidence that SC in CeAu$_{2}$Si$_{2}$ is mediated by critical spin fluctuations \cite{ReviewKnebel}. The fact that the onsets of filamentary and bulk SCs correlate with the anomalies in the $T_{\rm M}$ evolution (see also Fig. 2a), hinting at the existence of two putative QCPs, points to a deep link between SC and magnetic instabilities. However, the simultaneous enhancement of both SC and magnetic order in a wide volume (pressure) range, which has never been seen in any other Ce-based pressure-induced superconductors, can hardly be explained by this scenario. Instead, it is plausible that SC and magnetic order are not intrinsically related phenomena, although the possibility that SC develops from the magnetic ordered state cannot be excluded. This is further corroborated by the comparison of CeAu$_{2}$Si$_{2}$ and CeCu$_{2}$$X$$_{2}$, made in Fig. 3, which shows that a similar maximum $T_{\rm c}$ occurs regardless of the presence or the absence of a magnetic QCP. Thus, it appears that another pairing mechanism is involved at least on the low volume side of the superconducting region.

Another interesting clue pointing in this last direction is found in Fig. 3: for all three compounds, the maximum $T_{\rm c}$ occurs when the temperatures $T_{\rm 1}^{\rm max}$ and $T_{\rm 2}^{\rm max}$ of the resistivity maxima (as defined in Fig. 1a) merge, at relatively high temperature ($\sim$ 40$T_{\rm c}^{\rm max}$). Notice that in CeAu$_{2}$Si$_{2}$, $T_{\rm 1}^{\rm max}$ joins $T_{\rm 2}^{\rm max}$ at a slightly larger $V$ than that for CeCu$_{2}$$X$$_{2}$, similar to the small $V$-shift observed for the $T_{\rm c}$ maximum. Since the three quantities have been measured at each pressure run, their correspondence is unaffected by the uncertainty of pressure determination and hence significant. Although the exact relationships are yet to be determined, it is empirically known that the temperatures $T_{\rm 1}^{\rm max}$ and $T_{\rm 2}^{\rm max}$ scale approximately with the Kondo temperature ($T_{\rm K}$) and crystal field (CF) splitting energy, respectively \cite{twomaxima}. In our case, $T_{\rm 1}^{\rm max}$ gives an indication of $T_{\rm K}$ only for $V$ $<$ 165 {\AA}$^{3}$ ($i.e.$ $T_{\rm 1}^{\rm max}$ $>$ 10 K) when the low temperature resistivity maximum is free from the influence of magnetic ordering. As can be seen in Fig. 3, for $V$ $<$ 165 {\AA}$^{3}$, $T_{\rm 1}^{\rm max}$ ($\propto$ $T_{\rm K}$) shows a nearly exponential increase with decreasing $V$ and appears as the driving parameter of the system, which makes it evolve from long range magnetic ordered states, through SC, towards a strongly delocalized paramagnetic $f$-metal at reduced volume. Therefore, we conclude that the superconducting pairing is the strongest when Kondo and CF splitting energy scales become comparable. Moreover, $T_{\rm 1}^{\rm max}$ governs the ground state excitations reflected by the low temperature resistivity in the paramagnetic phase. For CeCu$_{2}$$X$$_{2}$ the relationship $T_{\rm 1}^{\rm max}$ $\propto$ 1/$\sqrt{A}$ (where $A$ is the Fermi liquid resistivity coefficient) was shown to be fulfilled, except around $p^{*}$ where $A$ abruptly drops by one order of magnitude \cite{CeCu2Si2Holmes}. Hence, above $p_{\rm c}$, the $A$ values of CeAu$_{2}$Si$_{2}$ (Inset of Fig. 2a) are similar to those of CeCu$_{2}$$X$$_{2}$ taken at the same $V$. Finally, we note that in comparison with CeCu$_{2}$$X$$_{2}$, $T_{\rm 2}^{\rm max}$ of CeAu$_{2}$Si$_{2}$ is almost two times higher, and $T_{\rm 1}^{\rm max}$ shows a slower rise for $V$ just below 165 {\AA}$^{3}$, which could account for the persistence of magnetism according to Doniach's simple scheme \cite{Doniach}. In passing, we remark that for CePd$_{2}$Si$_{2}$, SC also occurs when both Kondo resistivity and crystal field contribution peaks merge \cite{CePd2X2,CePd2Si2demuer}. It can be conjectured that this feature is a generic property of Ce-based HF superconductors.

In previous publications, the high-pressure superconducting dome of CeCu$_{2}$$X$$_{2}$ was interpreted within the framework of the critical valence fluctuation theory \cite{CeCu2Si2Holmes,Miyakireview,Holmesreview}. According to this approach, the critical end point of the valence transition line of the Ce ion lies at a pressure $p_{\rm v}$ and a temperature $T_{\rm cr}$ close to zero Kelvin. This theory provides a consistent interpretation of most of the features observed in the vicinity of $p_{\rm v}$, which include besides SC, a collapse of the resistivity associated with a $T$-linear regime, an enhanced residual resistivity and the above mentioned merging of the two temperatures $T_{\rm 1}^{\rm max}$ and $T_{\rm 2}^{\rm max}$ \cite{CeCu2Si2Holmes,ReviewJaccard2,Holmesreview}. However, around $p_{\rm v}$, calculations predict a strong decrease of the Ce-4$f$ orbital occupancy with increasing $p$, while X-ray absorption measurements show a considerably smaller variation (by a factor of 5) \cite{CeCu2Si2XAS}. This disagreement and the clue that at the maximum $T_{\rm c}$, the energy scale of the system is of the order of the CF splitting energy drew us to examine the role played by the orbital states and the associated fluctuations of the Ce-4$f$ electrons in the properties of these materials \cite{Hattori2010,Pourovskii2014}.

%% BEGIN THEORY DISCUSSION%%

\subsection{C. Comparison to dynamical mean-field theory calculations}
As a first step to address this issue, we performed calculations based on a combination of electronic structure
and dynamical mean-field theory (DMFT) methods,
along similar lines as in Ref. \cite{Pourovskii2014} for CeCu$_2$Si$_2$.
In a tetragonal crystal field, the $^2F_{5/2}$ ground-state multiplet of the Ce$^{3+}$ ion is split into three doublets:
$|0\rangle=a|\pm 5/2\rangle+\sqrt{1-a^2}|\mp 3/2\rangle$,
$|1\rangle=|\pm 1/2\rangle$,
and $|2\rangle=\sqrt{1-a^2}|\pm 5/2\rangle-a|\mp 3/2\rangle$.
A key difference between CeAu$_2$Si$_2$ and CeCu$_2$Ge$_2$ is already apparent
from our results calculated at ambient pressure and lowest pertinent temperature ($7$~K).
While for both compounds the ground-state is associated with state $|0\rangle$, the first excited
state is $|1\rangle$ in CeAu$_2$Si$_2$ but $|2\rangle$ in CeCu$_2$Ge$_2$.
\begin{figure}
\begin{center}
\includegraphics[width=8cm]{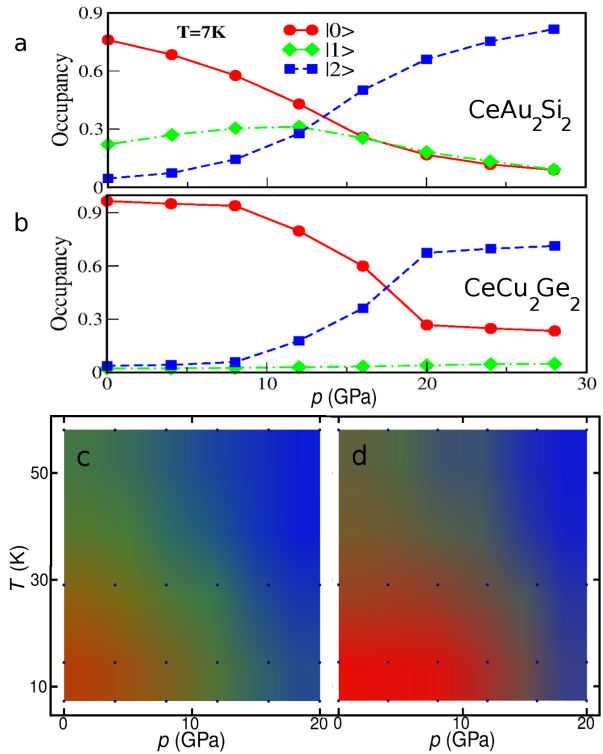}
\end{center}
\caption{(a) The calculated occupancy of the CF states $|0\rangle$ (circles), $|1\rangle$ (diamonds), and $|2\rangle$ (squares)
as a function of pressure at $T=$ 7 K for CeAu$_2$Si$_2$. The curves are linear interpolations between the corresponding points.
(b) The same data for CeCu$_2$Ge$_2$.
(c, d) The calculated $(T, p)$ maps of the orbital occupancies for CeAu$_2$Si$_2$ and CeCu$_2$Ge$_2$, respectively.
The color is defined by an RGB code in which the red, blue and green contributions are proportional to the occupancies $n_0$, $n_2$ and the sum of occupancies of two non-dominant states, respectively. Hence, the states $|0\rangle$ and $|2\rangle$ clearly dominate in the red and blue areas, respectively, while in the green region the occupancies of all three states are comparable.
The dots indicate the values of $T$ and $p$ for which the LDA+DMFT calculations were performed.
}
\end{figure}
%the energy difference between these two states is much smaller in
%CeAu$_2$Si$_2$ than in CeCu$_2$Ge$_2$.
%
In order to fully take the hybridization and Kondo effects into account, this splitting can be obtained from the
location of the respective Kondo peaks in the orbitally-resolved spectral
functions. For CeAu$_2$Si$_2$, we obtain levels $|1\rangle$ and $|2\rangle$ to be respectively 9.8 and 24.5 meV above the main Kondo peak associated with
state $|0\rangle$, in reasonable agreement with the reported experimental values of 17 and 21 meV~\cite{Willers_phd}.
By contrast, for CeCu$_2$Ge$_2$, we find these splitting to be 34  and 19 meV, respectively.
Note that for CeCu$_2$Ge$_2$ the ambient-pressure ground-state has been experimentally ascribed to state $|2\rangle$,
although this identification is based on simulations of the temperature dependence of the uniform magnetic susceptibility with the CF
levels treated as quasi-atomic levels neglecting hybridization and Kondo effects \cite{Knopp1989}.

Correspondingly, another key difference between CeCu$_2$Ge$_2$ and CeAu$_2$Si$_2$
is that the occupation of state $|1\rangle$ at ambient pressure is very small for the former while it is sizeable for the latter,
as displayed on Fig. 4 (upper panels).
In this figure, we plot the evolution of the occupancies of each state as a function of pressure for the two compounds.
For CeCu$_2$Ge$_2$, the occupancy of state $|1\rangle$ remains negligible at all pressures, but a transition
between a regime dominated by state $|0\rangle$ at low pressure and a regime dominated by state $|2\rangle$ at
high pressure takes place, with the occupancy of the two levels crossing each other around $17$~GPa. Across the transition region,
the $f$ electron weight is transferred from the CF state $|0\rangle$ to the excited level $|2\rangle$, due to the latter's stronger hybridization with itinerant electrons.
This `orbital transition' is quite similar to the one recently discussed theoretically~\cite{Hattori2010,Pourovskii2014} for CeCu$_2$Si$_2$, except that
it is shifted to higher pressure by about $15$~GPa in the Ge-based compound.
In contrast, the pressure evolution in the upper panel of Fig. 4 clearly displays {\it three} distinct regimes:
one dominated by state $|0\rangle$ at low pressure (roughly below $10$~GPa),
one dominated  by state $|2\rangle$ at high pressure ($\gtrsim 20$~GPa),
and an additional intermediate regime (roughly between $10$ and $20$~GPa)
where all three states contribute.
We have also followed the evolution of these three regimes as a function of temperature, and the
result is visualized in Fig. 4c and d as a color map in the $(p,T)$ plane.

Although our calculations are not performed in the phase with magnetic long-range order,
our results do hint at a qualitatively different behavior of the two compounds, as observed
experimentally.
It is tempting in particular to relate the three different regimes found for CeAu$_2$Si$_2$
to the observed persistence and revival of magnetism in the $15-20$~GPa range, and
possibly to the existence of several different magnetic phases (as suggested by the
kinks and non-monotonous behavior of the magnetic transition temperature $T_{\rm M}$ {\it vs.} pressure).
In contrast, in CeCu$_2$Ge$_2$, one observes a single magnetic phase which collapses at a
significantly lower pressure than the maximum of the SC dome.
Moreover, the pressure evolution of the Kondo temperature $T_{\rm K}$ ($\propto$ $T_1^{max}$)
in CeAu$_2$Si$_2$ (Fig.~\ref{fig3}) is different than in CeCu$_2$Ge$_2$, with an
intermediate slower increase in the range $10-16$~GPa. Indeed, this is consistent with our
calculated evolution of the effective mass, which displays a slow
decrease in the intermediate regime, followed by a faster one when state $|2\rangle$
dominates (Fig. S6 of the Supplemental Material \cite{SM}).

\section{IV. Summary and conclusion}
The above results underline the role of orbital physics in CeAu$_2$Si$_2$ and CeCu$_2$$X$$_2$. The critical end point, identified at ($p_{\rm cr}$, $T_{\rm cr}$) through the scaling of the resistivity around $p^{*}$ in CeAu$_2$Si$_2$, as well as previously in CeCu$_2$Si$_2$ \cite{CeCu2Si2seyfarth}, can be that of an orbital transition line, first described by Hattori \cite{Hattori2010} and in agreement with our calculations (see also Ref. \cite{Pourovskii2014}). In these systems, only a crossover regime is realized because the temperature $T_{\rm cr}$ is slightly negative. However, $T_{\rm cr}$ is small enough that the charge or orbital fluctuations associated to the orbital crossover are sufficiently developed to mediate both the non-Fermi liquid properties of the normal phase and the superconducting pairing. In CeAu$_{2}$Si$_{2}$, magnetism and superconductivity may originate from the occupancy of different CF levels in the intermediate pressure region according to our calculations. The increase of Kondo scale induced by the orbital transition may drive the collapse of magnetism, explaining its sudden disappearance and hence the proximity of the pressures $p_{\rm c}$ and $p^{*}$, as opposed to the case of CeCu$_{2}$$X$$_{2}$ for which these two pressures are well separated. As discussed in Sec. III B, it appears that spin fluctuations are not the driving force for superconductivity in CeAu$_{2}$Si$_{2}$. Anyway, it remains challenging to understand the giant overlap of SC and magnetism, and in particular the striking relationship $T_{\rm c}$ $\propto$ $T_{\rm M}$ observed in a broad pressure range, which definitely require further studies.

In conclusion, CeAu$_2$Si$_2$ has been discovered to be a new HF superconductor under a very broad pressure interval from 11.8 to 26.6 GPa. Within approximately two-thirds of this interval, SC appears below the magnetic phase transition. Intriguingly, when increasing pressure from 16.7 to 20.2 GPa both bulk $T_{\rm c}$ and $T_{\rm M}$ are strongly enhanced, and almost proportional. $T_{\rm c}$ reaches its maximum value of $\sim$ 2.5 K slightly below the pressure $p_{\rm c}$ $\thickapprox$ 22.5 GPa, where magnetic order disappears. The scaling behavior of resistivity indicates a continuous delocalization of Ce 4$f$ electrons associated with a critical end point lying just above $p_{\rm c}$. The $T_{\rm c}$ maximum occurs when the Kondo and CF energies are similar and at almost the same unit-cell volume as for CeCu$_{2}$Si$_{2}$ and CeCu$_{2}$Ge$_{2}$, providing a clue to the pairing mechanism. First-principle calculations indicate the existence of an "intermediate" state in the Ce 4$f$ orbital occupancy in CeAu$_{2}$Si$_{2}$, which might be related to its peculiar behavior in comparison with its close relatives CeCu$_{2}$$X_{2}$. Nevertheless, we emphasize that the understanding of the newly observed behavior in CeAu$_2$Si$_2$ remains largely open \cite{Flint}. Future experimental investigations of the isoelectronic compounds CeAg$_{2}$Si$_{2}$, or even CeAu$_{2}$Ge$_{2}$ and CeAg$_{2}$Ge$_{2}$, will likely enrich the debate. On the theoretical side, calculations of various Ce-based systems are highly desirable in order to extend comparisons with the already rich experimental results. For example, an interesting issue is the very weak pressure response of the "intermediate valence" compound CePd$_{3}$ \cite{CePd3}.

\section{ACKNOWLEDGEMENT}
\begin{acknowledgments}
We acknowledge discussions with M.~Ferrero, P.~Hansmann and J.-P.~Rueff, technical assistance from M. Lopes, and financial support from the Swiss National Science Foundation.
Computing resources were provided by the Swedish National Infrastructure for Computing (SNIC)
at the National Supercomputer Centre (NSC) and PDC Center for High Performance Computing,
and the Swiss Center for Scientific Computing.
\end{acknowledgments}


\begin{thebibliography}{99}
\expandafter\ifx\csname natexlab\endcsname\relax\def\natexlab#1{#1}\fi
\expandafter\ifx\csname bibnamefont\endcsname\relax
  \def\bibnamefont#1{#1}\fi
\expandafter\ifx\csname bibfnamefont\endcsname\relax
  \def\bibfnamefont#1{#1}\fi
\expandafter\ifx\csname citenamefont\endcsname\relax
  \def\citenamefont#1{#1}\fi
\expandafter\ifx\csname url\endcsname\relax
  \def\url#1{\texttt{#1}}\fi
\expandafter\ifx\csname urlprefix\endcsname\relax\def\urlprefix{URL }\fi
\providecommand{\bibinfo}[2]{#2}
\providecommand{\eprint}[2][]{\url{#2}}


\bibitem{CeCu2Ge2Jaccard}
D. Jaccard, K. Behnia and J. Sierro, \emph{Pressure induced heavy fermion superconductivity in} CeCu$_{2}$Ge$_{2}$, Phys. Lett. A {\bf 163}, 475 (1992).

\bibitem{CePd2Si2}
F. M. Grosche, S. R. Julian, N. D. Mathur, and G. G. Lonzarich, \emph{Magnetic and superconducting phases of} CePd$_{2}$Si$_{2}$,
Physica B {\bf 223-224}, 50 (1996).

\bibitem{CeIn3}
N. D. Mathur, F. M. Grosche, S. R. Julian, I. R. Walker, D. M. Freye, R. K. W. Haselwimmer and G. G. Lonzarich, \emph{Magnetically mediated superconductivity in heavy fermion compounds}, Nature {\bf 394}, 39 (1998).

\bibitem{CeRhIn5}
H. Hegger, C. Petrovic, E. G. Moshopoulou, M. F. Hundley, J. L. Sarrao, Z. Fisk, and J. D. Thompson, \emph{Pressure-induced superconductivity in quasi-2D} CeRhIn$_{5}$, Phys. Rev. Lett. {\bf 84}, 4986 (2000).

\bibitem{CePt2In7}
E. D. Bauer, H. O. Lee, V. A. Sidorov, N. Kurita, K. Gofryk, J.-X. Zhu, F. Ronning, R. Movshovich, J. D. Thompson, and T. Park, \emph{Pressure-induced superconducting state and effective mass enhancement near the antiferromagnetic quantum critical point of} CePt$_{2}$In$_{7}$, Phys. Rev. B {\bf 81}, 180507 (R) (2010).

\bibitem{ReviewKnebel}
G. Knebel, D. Aoki, and J. Flouquet, \emph{ Antiferromagnetism and superconductivity in cerium based heavy-fermion compounds}, Comptes Rendus Physique {\bf 12}, 542 (2011).

\bibitem{CeCu2Si2steglich}
F. Steglich, J. Aarts, C. D. Bredl, W. Lieke, D. Meschede, W. Franz, and H. Schafer, \emph{Superconductivity in the presence of strong Pauli paramagnetism}: CeCu$_{2}$Si$_{2}$, Phys. Rev. Lett. {\bf 43}, 1892 (1979).

\bibitem{CeCu2Si284}
B. Bellarbi, A. Benoit, D. Jaccard, J. M. Mignot, and H. F. Braun, \emph{High-pressure valence instability and $T_{c}$ maxium in superconducting} CeCu$_{2}$Si$_{2}$, Phys. Rev. B {\bf 30}, 1182 (1984).

\bibitem{ReviewDidier}
D. Jaccard, H. Wilhelm, K. Alami-Yadri, and E. Vargoz, \emph{Magnetism and superconductivity in heavy fermion compounds at high pressure}, Physica B {\bf 259-261}, 1 (1999).

\bibitem{CeCu2Si2Holmes}
A. T. Holmes, D. Jaccard and K. Miyake, \emph{Signatures of valence fluctuations in} CeCu$_{2}$Si$_{2}$ \emph{under high pressure}, Phys. Rev. B {\bf 69}, 024508 (2004).

\bibitem{ReviewJaccard2}
D. Jaccard and A. T. Holmes, \emph{Spin and valence-fluctuation mediated superconductivity in pressurized Fe and} CeCu$_{2}$(Si/Ge)$_{2}$, Physica B {\bf 359-361}, 333 (2005).

\bibitem{CeCu2Si2seyfarth}
G. Seyfarth, A.-S. Ruetschi, K. Sengupta, A. Georges, D. Jaccard, S. Watanabe, and K. Miyake, \emph{Heavy fermion superconductor} CeCu$_{2}$Si$_{2}$ \emph{under high pressure: multiprobing the valence crossover}, Phys. Rev. B {\bf 85}, 205105 (2012).

\bibitem{vargoz}
E. Vargoz and D. Jaccard, \emph{Superconducting and normal properties of} CeCu$_{2}$Ge$_{2}$ \emph{at high pressure},
J. Magn. Magn. Mater. {\bf 177-181}, 294 (1998).


\bibitem{CeCu2Si2yuan}
H. Q. Yuan, F. M. Grosche, M. Deppe, C. Geibel, G. Sparn, and F. Steglich \emph{Observation of two distinct superconducting phases in} CeCu$_{2}$Si$_{2}$, Science {\bf 302}, 2104 (2003).

\bibitem{CeCu2Si2stockert}
O. Stockert, J. Arndt, E. Faulhaber, C. Geibel, H. S. Jeevan, S. Kirchner, M. Loewenhaupt, K. Schmalzl, W. Schmidt, Q. Si and F. Steglich, \emph{Magnetically driven superconductivity in} CeCu$_{2}$Si$_{2}$, Nat. Phys. {\bf 7}, 119 (2011).

\bibitem{Lonzarich}
P. Monthoux, D. Pines and G. G. Lonzarich, \emph{Superconductivity without phonons}, Nature {\bf 450}, 1177 (2007).

\bibitem{CeCu2Si2multiband}
S. Kittaka, Y. Aoki, Y. Shimura, T. Sakakibara, S. Seiro, C. Geibel, F. Steglich, H. Ikeda, and K. Machida, \emph{Multiband superconductivity with unexpected deficiency of nodal quasiparticles in} CeCu$_{2}$Si$_{2}$, Phys. Rev. Lett. {\bf 112}, 067002 (2014).

\bibitem{CeCu2Si2XAS}
J.-P. Rueff, S. Raymond, M. Taguchi, M. Sikora, J.-P. Itie, F. Baudelet, D. Braithwaite, G. Knebel, and D. Jaccard, \emph{Pressure-induced valence crossover in superconducting} CeCu$_{2}$Si$_{2}$, Phys. Rev. Lett. {\bf 106}, 186405 (2011).

\bibitem[{\citenamefont{Hattori}(2010)}]{Hattori2010}
K. Hattori, \emph{Meta-orbital transition in heavy-fermion systems: analysis by dynamical mean field theory and self-consistent renormalization theory of orbital fluctuations}, J. Phys. Soc. Jpn. {\bf 79}, 114717 (2010).

\bibitem[{\citenamefont{Pourovskii et~al.}(2014)\citenamefont{Pourovskii,
  Hansmann, Ferrero, and Georges}}]{Pourovskii2014}
L. V. Pourovskii, P. Hansmann, M. Ferrero and A. Georges, \emph{Theoretical prediction and spectroscopic fingerprints of an orbital transition in} CeCu$_{2}$Si$_{2}$, Phys. Rev. Lett. {\bf 112}, 106407 (2014).


\bibitem{CeAu2Si2Link}
P. Link and D. Jaccard, \emph{Pressure induced heavy-fermion behavior of} CeAu$_{2}$Si$_{2}$ \emph{near 17 GPa}, Physica B {\bf 230-232}, 31 (1997).

\bibitem{canfield}
P. C. Canfield and Z. Fisk, \emph{Growth of single crystals from metallic fluxes},
Philos. Mag. B {\bf 65}, 1117 (1992).

\bibitem{elementCe}
Materials Preparation Center, Ames Laboratory, US DOE Basic Energy Sciences, Ames, IA, USA, available from http://www.mpc.ameslab.gov.

\bibitem{CeAu2Si2crystalgrowth}
Y. Ota, K. Sugiyama, Y. Miyauchi, Y. Takeda, Y. Nakano, Y. Doi, K. Katayama, N. D. Dung, T. D. Matsuda, Y. Haga, K. Kindo, T. Takeuchi, M. Hagiwara, R. Settai, and Y. Onuki, \emph{Electrical and magnetic properties of} CeAu$_{2}$Si$_{2}$, J. Phys. Soc. Jpn. {\bf 78}, 034714 (2009).

\bibitem{Holmesiron}
A. T. Holmes, D. Jaccard, G. Behr, Y. Inada, and Y. Onuki, \emph{Unconventional superconductivity and non-Fermi liquid behavior of $\epsilon$-iron at high pressure}, J. Phys.:Condens. Matter {\bf 16}, S1121 (2004).

\bibitem{SM}
See Supplemental Material at [\emph{URL will be inserted by publisher}] for more data and analysis.

\bibitem{Holmesthesis}
A. T. Holmes, \emph{Exotic superconducting mechanism in Fe and} CeCu$_{2}$Si$_{2}$ \emph{under pressure}, Ph.D. thesis, Universit$\acute{e}$ de Gen$\grave{e}$ve (2004), http://archive-ouverte.unige.ch/unige:284.

\bibitem[{\citenamefont{Aichhorn et~al.}(2009)\citenamefont{Aichhorn,
  Pourovskii, Vildosola, Ferrero, Parcollet, Miyake, Georges, and
  Biermann}}]{Aichhorn2009}
M. Aichhorn, L. Pourovskii, V. Vildosola, M. Ferrero, O. Parcollet, T. Miyake, A. Geroges, and S. Biermann, \emph{Dynamical mean-field theory within an augmented plane-wave framework: assessing electronic correlations in the iron pnictide LaFeAsO}, Phys. Rev. B {\bf 80}, 085101 (2009).

\bibitem[{\citenamefont{Aichhorn et~al.}(2011)\citenamefont{Aichhorn,
  Pourovskii, and Georges}}]{Aichhorn2011}
M. Aichhorn, L. Pourovskii and A. Georges, \emph{Importance of electronic correlations for structural and magnetic properties of the iron pnictide superconductor LaFeAsO},
Phys. Rev. B {\bf 84}, 054529 (2011).

\bibitem[{\citenamefont{Blaha et~al.}(2001)\citenamefont{Blaha, Schwarz,
  Madsen, Kvasnicka, and Luitz}}]{Wien2k}
P. Blaha, K. Schwarz, G. Madsen, D. Kvasnicka, and J. Luitz,
  \bibinfo{title}{\emph{WIEN2k, An augmented plane wave + local orbitals
  program for calculating crystal properties}} (\bibinfo{publisher}{Techn.
  Universit$\ddot{a}$t Wien, Austria}, \bibinfo{year}{2001}).

\bibitem[{\citenamefont{Gull et~al.}(2011)\citenamefont{Gull, Millis,
  Lichtenstein, Rubtsov, Troyer, and Werner}}]{Gull2011}
E. Gull, A. J. Millis, A. I. Lichtenstein, A. N. Rubtsov, M. Troyer, and P. Werner, \emph{Continuous-time Monte Carlo methods for quantum impurity models}, Rev. Mod. Phys. {\bf 83}, 349 (2011).

\bibitem[{\citenamefont{Ferrero and Parcollet}()}]{TRIQS}
M. Ferrero and O. Parcollet,
  \bibinfo{note}{\emph{TRIQS: a toolbox for research on interacting quantum
  systems, http://ipht.cea.fr/triqs}}.

\bibitem{LDAestimation}
Our estimates within the constrained
LDA method for CeCu$_2$Si$_2$, CeCu$_2$Ge$_2$, and CeAu$_2$Si$_2$ give
values of  $F_0$ differing by less than 0.2~eV. Hence, the value
previously used for  CeCu$_2$Si$_2$ has been adopted for the two other
compounds.

\bibitem[{\citenamefont{Onodera et~al.}(2002)\citenamefont{Onodera, Tsuduki,
  Ohishi, Watanuki, Ishida, Kitaoka, and Onuki}}]{Onodera2002}
A. Onodera, S. Tsuduki, Y. Ohishi, T. Watanuki, K. Ishida, Y. Kitaoka, and Y. Onuki, \emph{Equation of state of} CeCu$_{2}$Ge$_{2}$ \emph{at cryogenic temperature},
Solid State Commun. {\bf 123}, 113 (2002).

\bibitem{CeCu2Ge2latticeparameter}
S. Tsuduki, A. Onodera, K. Ishida, Y. Kitaoka, A. Onuki, N. Ishimatsu, and O. Shimomura, \emph{Synchrotron X-ray diffraction and absorption studies of} CeM$_{2}$X$_{2}$ (M=Cu, Ni and X=Si, Ge) \emph{at high pressure}, Solid State Commun. {\bf 134}, 747 (2005).

\bibitem{CeAu2Si2latticeparameter}
M. Ohmura, K. Sakai, T. Nakano, H. Miyagawa, G. Oomi, I. Sato, T. Komatsubara, H. Aoki, Y. Matsumoto, and Y. Uwatoko, \emph{Anisotropic lattice compression and possible valence change in Kondo compound} CeAu$_{2}$Si$_{2}$, J. Magn. Soc. Jpn. {\bf 33}, 31 (2009).

\bibitem{phononcontribution}
The phonon contribution $\rho_{\rm ph}$ is well approximated by a linear in-$T$ term above $\Theta_{\rm D}$/10, where $\Theta_{\rm D}$ is the Debye temperature. From specific heat data of Ref. 24, $\Theta_{\rm D}$ $\approx$ 270 K is obtained. Below $\Theta_{\rm D}$/10 $\approx$ 27 K, the linear term overestimates $\rho_{\rm ph}$. However, the low $T$ resistivity is dominated by the magnetic contribution even at low pressure, and thus the subtraction of a small linear term has weak effect on our data analysis. Under pressure, $\rho_{\rm ph}$ is expected to decrease slightly, but in comparison to the strong increase of the magnetic resistivity, it can be taken as pressure independent.

\bibitem{scattering1}
B. Cornut and B. Coqblin, \emph{Influence of the crystalline field on the Kondo effect of alloys and compounds with cerium impurities}, Phys. Rev. B {\bf 5}, 4541 (1972).

\bibitem{texturedSC}
T. Park, H. Lee, I. Martin, X. Lu, V. A. Sidorov, K. Gofryk, F. Ronning, E. D. Bauer, and J. D. Thompson, \emph{Textured superconducting phase in the heavy fermion} CeRhIn$_{5}$, Phys. Rev. Lett. {\bf 108}, 077003 (2012).

\bibitem{determinationofTM}
N. Tateiwa, Y. Haga, T. D. Matsuda, S. Ikeda, T. Yasuda, T. Takeuchi, R. Settai, and Y. Onuki,
\emph{Novel pressure phase diagram of heavy fermion superconductor} CePt$_{3}$Si \emph{investigated by ac calorimetry},
J. Phys. Soc. Jpn. {\bf 74}, 1903 (2005).


\bibitem{CeAuSb2}
S. Seo, V. A. Sidorov, H. Lee, D. Jang, Z. Fisk, J. D. Thompson, and T. Park, \emph{Pressure effects on the heavy-fermion antiferromagnet} CeAuSb$_{2}$, Phys. Rev. B {\bf 85}, 205145 (2012).

\bibitem{Hc2slope}
T. P. Orland, E. J. McNiff, Jr., S. Foner, and M. R. Beasley, \emph{Critical fields, Pauli paramagnetic limiting, and material parameters of} Nb$_{3}$Sn \emph{and} V$_{3}$Si. Phys. Rev. B {\bf 19}, 4545 (1979).

\bibitem{CeCuSi2NQR}
T. C. Kobayashi, K. Fujiwara, K. Takeda, H. Harima, Y. Ikeda, T. Adachi, Y. Ohishi, C. Geibel, and F. Steglich, \emph{Valence crossover of Ce ions in} CeCu$_{2}$Si$_{2}$ \emph{under high pressure --Pressure dependence of the unit cell volume and the NQR frequency--}, J. Phys. Soc. Jpn. {\bf 82}, 114701 (2013).

\bibitem{CePd2X2}
H. Wilhelm and D. Jaccard, \emph{Calorimetric and transport investigations of} CePd$_{2+x}$Ge$_{2-x}$ \emph{($x$=0 and 0.02) up to 22 GPa},
Phys. Rev. B {\bf 66}, 064428 (2002).

\bibitem{twomaxima}
Y. Nishida, A. Tsuruta, and K. Miyake,
\emph{Crystalline-electric-field effect on the resistivity of Ce-based heavy fermion systems},
J. Phys. Soc. Jpn. {\bf 75}, 064706 (2006).

\bibitem{Doniach}
S. Doniach,
\emph{The Kondo lattice and weak antiferromagnetism},
Physica B {\bf 91}, 231 (1977).

\bibitem{CePd2Si2demuer}
A. Demuer, A. T. Holmes, and D. Jaccard, \emph{Strain enhancement of superconductivity in} CePd$_{2}$Si$_{2}$ \emph{under pressure},
J. Phys.: Condens. Matter {\bf 14}, L529 (2002).

\bibitem{Miyakireview}
K. Miyake, \emph{New trend of superconductivity in strongly correlated electron systems},
J. Phys.: Condens. Matter {\bf 19}, 125201 (2007).

\bibitem{Holmesreview}
A. T. Holmes, D. Jaccard and K. Miyake,
\emph{Valence instability and superconductivity in heavy fermion systems},
J. Phys. Soc. Jpn. {\bf 76}, 051002 (2007).

\bibitem[{\citenamefont{Georges et~al.}(1996)\citenamefont{Georges, Kotliar,
  Krauth, and Rozenberg}}]{Georges1996}
A. Georges, G. Kotliar, W. Krauth, and M. J. Rozenberg, \emph{Dynamical mean-field theory of strongly correlated fermion systems and the limit of infinite dimensions}, Rev. Mod. Phys. {\bf 68}, 13 (1996).



\bibitem[{\citenamefont{Willers}(2011)}]{Willers_phd}
T. Willers, \bibinfo{title}{\emph{Spectroscopic investigations of the crystal field and
  Kondo effect in 4\emph{f} heavy-fermion systems}, PhD thesis},
  \bibinfo{publisher}{Universit\"at zu K\"oln} (\bibinfo{year}{2011}).

\bibitem[{\citenamefont{Knopp et~al.}(1989)\citenamefont{Knopp, Loidl, Knorr,
  Pawlak, Duczmal, Caspary, Gottwick, Spille, Steglich, and
  Murani}}]{Knopp1989}
G. Knopp, A. Loidl, K. Knorr, L. Pawlak, M. Duczmal, R. Caspary, U. Gottwick, H. Spille, F. Steglich, and A. P. Murani, \emph{Magnetic order in a Kondo lattice: a neutron scattering study of} CeCu$_{2}$Ge$_{2}$, Z. Phys. B {\bf 77}, 95 (1989).

\bibitem{Flint}
R. Flint, A. H. Nevidomskyy, and P. Coleman, \emph{Composite pairing in a mixed-valent two-channel Anderson model},
Phys. Rev. B {\bf 84}, 064514 (2011).

\bibitem{CePd3}
P. Pedrazzini, D. Jaccard, M. Deppe, C. Geibel, and J. G. Sereni, \emph{Multiprobe high-pressure experiments in} CePd$_{0.6}$Rh$_{0.4}$ \emph{and} CePd$_{3}$, Physica B {\bf 404}, 2898 (2009).

\end{thebibliography}
\end{document}